\documentclass[aps,pra,nofootinbib,preprint,amsmath,amssymb,floatfix]{revtex4}
\usepackage{color}
\usepackage{graphicx}

\begin{document}

\newcommand{\tc}{\textcolor}
\newcommand{\g}{blue}
\newcommand{\ve}{\varepsilon}
\title{Does the spacelike character of the Minkowski four-momentum show up in analog gravity?}         

\author{  Iver Brevik  }      

\affiliation{Department of Energy and Process Engineering, Norwegian University of Science and Technology, N-7491 Trondheim, Norway}

\begin{abstract}
In analog gravity the recent experiment of Drori {\it et al.} [Phys. Rev. Lett. {\bf 122}, 010404 (2019)] is impressive, as it shows how the emission of two Hawking quanta emitted in opposite directions lead to measurable consequences in the medium's rest system in a straightforward way. This result raises however the following problem: how can this experiment be explained in terms of classical electrodynamics? There must necessarily exist such an explanation (the experiment is after all classical); otherwise classical electrodynamics would be an incomplete theory. This is the main topic of the present paper. We propose that the measured effect is a demonstration of the spacelike character of the Minkowski four-momentum. Moreover, we extend the discussion by analyzing a Gedanken experiment (making use of the Kerr effect as a formal agency), to illustrate the transition from subluminal to superluminal phenomena in a straightforward way.  Finally, we emphasize the close relationship that exists between the spacelike Minkowski momentum and the anomalous Doppler effect.

\end{abstract}
\maketitle

\noindent {\it 1. Introduction.} - Due to the lack of observations of Hawking radiation   from astrophysical objects \cite{hawking74}, the possibility of observing analogous kinds of radiation from terrestrial systems is a topic that has quite naturally attracted considerable interest. One pioneering paper in this direction was that of Unruh  \cite{unruh81}, in which a sonic-generated analog of a black hole horizon was analyzed. The central point here was the behavior of sound waves in an oppositely moving fluid when the fluid velocity was larger than the velocity of sound. Later on, several studies of analog gravity have appeared, among them how light propagates in moving dispersive media \cite{leonhardt00}, and how shallow water waves propagate in an opposing current under critical conditions \cite{schutzhold02,rousseaux08}. We will not  here give an extensive review over the many  developments, but refer the reader to two volumes \cite{novello02,faccio13}, and also to the extensive survey over recent background literature  given in the recent paper of Liao {\it et al.} \cite{liao19}.

The present note was motivated by the recent experimental and theoretical work of Drori {\it et al.} \cite{drori19}. In the experiment an artificial event horizon was created in the form of short light pulses propagating in a nonlinear medium, each pulse giving rise to a perturbation $\delta n$ of the refractive index via the Kerr effect. The velocity $u$ of the the pulse  was therewith a little lower than the value $c/n$ of
light in the nonperturbed medium in the laboratory frame $S$ (medium at rest). In the comoving frame of reference $S^\prime$ the pulse stands still,  the medium moving in the opposite direction with velocity $-u$. In the presence of probe light, the pulse stablishes at its leading edge a black hole horizon where $u$ becomes equal to the group velocity $c/(n+\omega dn/d\omega)$ of the probe.

Assume now that a pair of Hawking quanta of frequency $\pm \omega'$ are emitted in the frame $S^\prime$. The relationship between $\omega'$ and $\omega$ in the laboratory frame $S$ is given by the Doppler formula
\begin{equation}
\omega'=\gamma ( 1-nu/c)\omega, \label{1}
\end{equation}
with $\gamma=(1-u^2/c^2)^{-1/2}$.
By taking into account the dispersive properties of the medium, Drori {\it et al.} were able to relate the two frequencies $\pm \omega'$ to two {\it positive} frequencies in the laboratory frame, and verified their existence experimentally to a good accuracy.

These experimental result are quite impressive. The following question arises however immediately: the experiment is basically  classical in nature, and should as such be explainable in terms of classical electrodynamics only. If that were not the case, classical electrodynamics would simply be an incomplete theory. And this brings us to the first point: what is the classical explanation of these experimental findings? As we will argue in the next section, it is the {\it spacelike} character of the Minkowski four-momentum of a radiation field that seems to be the chief agency behind this  observable effect. The spacelike character of the electromagnetic four-momentum (Minkowski four-momentum) for a source-free field  does not show up frequently, but this case seems to be one of those rare cases. To our knowledge, this kind of explanation has not appeared before.

\bigskip
\noindent {\it 2.  Covariant electrodynamics applied to the experiment.} - We start from a general perspective, first establishing the notation. Following   Refs. \cite{moller72} or \cite{brevik79}, we write the four-coordinate as $x_\mu=({\bf r}, x_4)$ with $x_4=ix_0=ict$. Correspondingly, in Fourier space $k_\mu=({\bf k}, k_4)$ with $k_4=ik_0 =i\omega/c$. The four-velocity of the uniformly moving medium is $V_\mu=({\bf V}, V_4)$ with $V_4=iV_0$, satisfying $V_\mu V_\mu=-c^2$ or $V_0^2={\bf V}^2+c^2$, where $V_\mu=\gamma ({\bf v}, ic),~\gamma=(1-v^2/c^2)^{-1/2}$, ${\bf v}$ being the three-dimensional velocity of the medium.
There are in any inertial frame two field tensors, $F_{\mu \nu}$ and $H_{\mu \nu}$, related to the electric and magnetic fields via
$F_{ik}=B_l, \,  F_{4k}=(i/c)E_k, \, H_{ik}=H_l, \quad H_{4k}=ic D_k, \, (i,k,l$ cyclic). SI units are assumed.
Minkowski's energy-momentum tensor reads in covariant form,
\begin{equation}
S_{\mu\nu}^M=F_{\mu\alpha}H_{\nu\alpha}-\frac{1}{4}\delta_{\mu\nu} F_{\alpha\beta}H_{\alpha_\beta}.
\end{equation}.
Expressed in terms of the fields, we have for the spatial components
\begin{equation}
S_{ik}^M= -E_i D_k-H_i B_k+\frac{1}{2}\delta_{ik}{\bf (E\cdot D+H\cdot B)}, \label{a}
\end{equation}
and  Poynting's vector
${\bf S}^M$ and the momentum density ${\bf g}^M$ are
\begin{equation}
{\bf S}^M = {\bf E\times H}, \quad {\bf g}^M={\bf D\times B}. \label{b}
\end{equation}
Note that also the expressions (\ref{a}) and (\ref{b}) are covariant, holding in all inertial frames.

The constitutive relations in the rest frame of the medium will be written as
${\bf D}=\varepsilon {\bf E}$ with $\varepsilon=\varepsilon_0\varepsilon_r$ and ${\bf B}=\mu {\bf H}$ with $\mu=\mu_0\mu_r$. They can be expressed covariantly as
\begin{equation}
F_{\mu\nu}V_\alpha+F_{ \nu\alpha}V_\mu+F_{\alpha \mu}V_\nu=\mu (H_{\mu\nu}V_\alpha +H_{\nu \alpha}V_\mu+H_{\alpha \mu}V_\nu), \label{c}
\end{equation}
or, in three-dimensional form which for us will be more convenient,
\begin{equation}
{\bf D}+\frac{1}{c^2}\left( {\bf v\times H}\right)  =  \varepsilon ({\bf E}+{\bf v\times B}), \label{5}
\end{equation}
\begin{equation}
{\bf B}+\frac{1}{c^2}\left( {\bf E\times v} \right)= \mu ({\bf H}+{\bf D\times v}). \label{6}
\end{equation}
We can now derive the dispersion relation for a monochromatic wave moving in this medium, by starting from the Lagrangian density $L=-(1/4)F_{\mu\nu}H_{\mu\nu}$,
from which we obtain
\begin{equation}
k^2-\kappa (k\cdot V)^2=0,  \quad \kappa =\frac{n^2-1}{c^2},
\end{equation}
with $ \quad  k^2=k_\alpha k_\alpha, ~k\cdot V=k_\alpha V_\alpha$, and $n^2=\varepsilon_r\mu_r$. The solution for $k_0=\omega/c$ is
\begin{equation}
k_0=\frac{\kappa V_0({\bf k\cdot V}) \pm \sqrt{(1+\kappa V_0^2){\bf k}^2-\kappa ({\bf k\cdot V})^2}}{1+\kappa V_0^2}, \label{10}
\end{equation}
which shows that for given values of $\bf k$ and $\bf V$ there are in general two values for $\omega$.

Having now established the general formalism we will change our convention
 a little, in order to conform with that of Ref.~\cite{drori19}: let  $S^\prime$ be the inertial frame where the velocity of the medium is $\bf V$.  We let $\bf V$ be directed along the negative $z$ axis. Moreover, we take $\bf k'$ also to lie along the $z$ axis, ${\bf k'}=k'_z{\bf e}_z$, but permit $k'_z$ at first to have either sign. The dispersion relation (\ref{10}) thus simplifies to
\begin{equation}
k_0'= \frac{-\kappa V_0|{\bf V|} \pm n}{1+\kappa V_0^2}\,k_z'. \label{11}
\end{equation}
 We note that the nominator in Eq.~(\ref{10}) is zero in the luminal case
\begin{equation}
|\beta| \equiv \frac{|v_z|}{c}=\frac{1}{n}. \label{12}
\end{equation}
The relativistic transformations between $k_z', \omega'$ in $S^\prime$ and $k_z, \omega$ in the laboratory frame $S$ (where ${\bf k}=(n\omega/c)\hat{\bf k})$ are in the present notation
\begin{equation}
k_z'=(1-|\beta|/n)k_z, \label{13}
\end{equation}
\begin{equation}
\omega'= \gamma (1-n|\beta|)\omega. \label{14}
\end{equation}
We shall make the requirement that  all physically propagating  waves in the subluminal case $n|\beta| <1$ can be assigned with a positive frequency. That is always possible in classical electrodynamics, in any inertial frame. It implies that different  directions of propagation, such as in plus or minus directions along are given axis,  are taken care of via the wave vector, not via the sign of the frequency.

Now return to the experiment \cite{drori19}. Characteristically there were {\it two} different frequencies observed, both propagating in the same positive $z$ direction in the laboratory frame. That means, $k_z >0$ in both cases. As Eq.~(\ref{13}) shows,  $k_z'$ has then to be positive (one cannot reverse the direction of the wave vector by a Lorentz transformation). As we require $k_0'=\omega'/c$ to be positive, this means that only {\it one} of the two possibilities in Eq.~(\ref{11}) are realizable,
\begin{equation}
k_0'= \frac{-\kappa V_0|{\bf V|} + n}{1+\kappa V_0^2}\,k_z'. \label{15}
\end{equation}
This accounts for one of the observed waves, but where  does the second wave come from? In our opinion the most natural explanation is that it is the {\it spacelike property of the electromagnetic Minkowski four-momentum} that shows itself up here. When the medium moves superluminally, $n|\beta|>1$, the electromagnetic energy may become negative. As seen from Eq.~(\ref{15}), in the superluminal case the nominator becomes negative, and there is no problem in associating a negative value of $k_z'$ with a positive value of $k_0'$. Here we have made  use of the important property of the Minkowski theory that $\hbar k_\mu$ is just equal to the photon four-momentum. It is rather remarkable that the Minkowski theory fits so well into canonical field theory, as has been discussed earlier at various places, for instance in some detail in Ref.~\cite{brevik17}.

We have thus obtained a   classical counterpart to the reasoning  made in Ref.~\cite{drori19}, in that case based upon  a quantum mechanical viewpoint. Since the "particle" (the disturbance $\delta n$) moves very closely to the luminal case $c/n$ (a little below it because of the Kerr effect), it is feasible that the superluminal effect may turn up. Both classically and quantum mechanically, the two photons in the frame $S^\prime$ are emitted in opposite directions.

In general, the  spacelike property of Minkowski's four-momentum may be traced back to the fact that Minkowski's theory breaks the so-called Planck's principle of inertia of energy, ${\bf g}=(1/c^2)\bf S$, where $\bf g$ is the field momentum density and $\bf S$ the Poynting vector, and replaces it with the form ${\bf g}^M=(n^2/c^2)\bf S$. Minkowski's energy-momentum tensor is divergence-free for a homogeneous dielectric, thus making the total energy and momenta components form a four-vector, what  turns out to be most convenient for a quantum mechanical formulation, as noted already above. It is not so often that the  spacelike character of the Minkowski four-momentum turns out; a typical  example being the anomalous Doppler effect (see below). For an introduction to the Minkowski theory, one may consult M{\o}ller's book \cite{moller72}. Also, the present author has dealt with this tensor under various occasions \cite{brevik79,brevik17,brevik18}.

Before leaving this experiment, we ought to recall the importance of dispersion in the explanation given in \cite{drori19}. This implies, via the Kramers-Kronig relation and the fluctuation-dissipation theorem, that the retarded Green function has to possess a nonvanishing imaginary part (otherwise, the two-point functions would be zero). Thus, in analog gravity, as in quantum mechanics in general, some information must be lost to have any information at all. This remarkable property was recently  emphasized also by Hartle \cite{hartle19}.

One may ask: can one use the Abraham tensor to analyze this experiment? In principle one may do so, although the description becomes more cumbersome. The four-vector property of $k_\mu$ is maintained, but what is lost is the analogous property of radiation energy and momentum for the total field. The reason, again, is that the Abraham tensor is not divergence-free in a homogeneous medium. The close association between $\hbar k_\mu$ and photon momentum is therewith lost. Again, we may refer to our earlier treatment in  Ref.~\cite{brevik17}.

\bigskip

\noindent {\it 3. A Gedanken experiment.} - Related to our considerations above we  propose the following Gedanken experiment, designed such as to make the physics of the  subluminal-superluminal transition more transparent. Assume, in the inertial frame $S$, that there is a uniform flow with velocity $\bf v$ in the negative $z$ direction. We assume the magnitude $|\bf v|$ to lie slightly below the luminal limit $c/n$, where $n=\sqrt{\varepsilon_r\mu_r} >1$.  Thus $n|\beta| <1$, with $\beta=v/c$.  In such a fluid,  it will be possible for an electromagnetic wave to move from left to right. In the region $z>0$ we assume however that there is a strong transverse electric field, producing a slightly higher refractive index  $n_1$ in this region via  the Kerr effect.  Moreover, an important point    is  that the region $z>0$ is required to be superluminal, $n_1|\beta| >1$. We will analyze how an incident monochromatic wave from the left behaves near the divide $z=0$.

Assume that the incident field $E_I$ in the left region $z<0$ is polarized in the $x$ direction,
\begin{equation}
E_x=E_Ie^{i(k_Iz-\omega t)}, \label{16}
\end{equation}
where $k_I=k_z= \tilde{n}_I\omega/c$, $n_I$ being the effective refractive index,
\begin{equation}
\tilde{n}_I=\frac{1+\kappa V_0^2}{n-\kappa V_0|\bf V|}; \label{18}
\end{equation}
cf. Eq.~(\ref{15}) and the absence of primes with the present convention. From Maxwell's equations $\bf{\nabla \times E}=-\partial{\bf B}/\partial _Tt$, $\bf{\nabla \times H}=\partial {\bf D}/\partial t$, valid in any frame,  we have  $B_y=({\tilde n}_I/ c)E_I, \, H_y=(c/{\tilde n}_I) D_x$. To relate the latter fields to $E_I$, we can make use of the $y$ component of the constitutive relation (\ref{6}) to get
\begin{equation}
H_I=\frac{c}{{\tilde n}_I}D_I=\frac{E_I}{\mu c}\,\frac{\tilde{n}_I+|\beta|}{1+\tilde{n}_I|\beta|}. \label{20}
\end{equation}
This leads to the following energy density for the incident field (assuming $E_I$ to be real),
\begin{equation}
W_I=\frac{1}{4}\Re ({\bf {E\cdot D^*}}+ {\bf{H\cdot B^*}})= \frac{E_I^2}{2c^2}\, \frac{{\tilde n}_I}{\mu}\, \frac{{\tilde{n}}_I+|\beta|}{1+{\tilde{n}}_I|\beta|}. \label{21}
\end{equation}
The electric and magnetic fields contribute to this expression equally. For the incident Poynting vector we obtain similarly
\begin{equation}
S_I=\frac{1}{2}\Re ({\bf E\times H^*})_z= \frac{E_I^2}{2\mu c}\, \frac{{\tilde{n}}_I+|\beta|}{1+{\tilde{n}}_I|\beta|}. \label{22}
\end{equation}
We thus see that $S_I=W_I(c/{\tilde{ n}}_I)$, as we should expect since ${\tilde{n}}_I$ plays the role of an effective refractive index.

Next consider the reflected wave $E_Re^{i(k_Rz+\omega t)}$  in the region $z<0$. As $\bf k$ and $\bf V$ in Eq.~(\ref{10}) are now parallel, the wave vector is changed into $k_R=\tilde{n}_R\omega/c$, where the effective refractive index is
\begin{equation}
\tilde{n}_R=\frac{1+\kappa V_0^2}{n+\kappa V_0|{\bf V}|}. \label{23}
\end{equation}
As the reflection occurs at a resting surface $z=0$ in the frame $S$ the frequency $\omega$ is the same as the incident frequency (the surface force does no work).  The   reflected magnetic field along  the $y$ axis  is
\begin{equation}
H_R=\frac{E_R}{\mu c}\, \frac{{\tilde n}_R+|\beta|}{1+{\tilde n}_R |\beta|}, \label{24}
\end{equation}
(we assume a phase shift in $E_R$ relative to $E_I$ since the region $z>0$ is optically denser. This is however a formal point, and has no influence on the physical quantities).
In the transmitted region $z>0$ where the refractive index is $n_1$, the field is $E_T$ along the $x$ axis, and the effective refractive index is
\begin{equation}
\tilde{n}_T= \frac{1+\kappa_1V_0^2}{n_1-\kappa_1V_0|{\bf V}|} \label{25}
\end{equation}
with $\kappa_1=(n_1^2-1)/c^2$. The  magnetic field along the $y$ axis is
\begin{equation}
H_T= \frac{E_T}{\mu_1c}\, \frac{\tilde{n}_T+|\beta|}{1+\tilde{n}_T|\beta|}. \label{26}
\end{equation}
We will now require the usual continuity properties of the fields to hold at the surface $z=0$,
\begin{equation}
E_I-E_R=E_T, \quad H_I+H_R=H_T, \label{27}
\end{equation}
although one should observe that this point is not quite trivial:  strictly speaking there is in the present case no dielectric boundary  across which the material constants are known to vary continuously; cf. usual optics where this continuity property is known.

Inserting from Eqs.~(\ref{20}), (\ref{24}) and (\ref{26}) in Eqs.~(\ref{27}) we obtain the ratio between transmitted and incident  fields,
\begin{equation}
\frac{E_T}{E_I}= \frac{\mu_1}{\mu}\, \frac{(\tilde{n}_I+\tilde{n}_R)(1+\beta^2)+2(1+\tilde{n}_I\tilde{n}_R)|\beta|} {  (\tilde{n}_T+\tilde{n}_R)(1+\beta^2)+2(1+\tilde{n}_T\tilde{n}_R)|\beta|} \, \frac{1+{\tilde n}_T|\beta|}{1+\tilde{n}_I|\beta|}.
 \label{28}
\end{equation}
From this we can  calculate the transmitted quantities in terms of $E_I$:  Poynting vector,  momentum density, and  energy density. In the following we will however simplify the formalism using that the case of main physical interest is when the fluid velocity lies close to the luminal border $|\beta|=1/n$. Then we can insert $V_0=n/{\sqrt \kappa}, \, |{\bf V}|=1/{\sqrt{\kappa}}$, so that $\tilde{n}_I \rightarrow \infty,  \, \tilde{n}_R \rightarrow \frac{n^2+1}{2n}$. We obtain
\begin{equation}
S_I= \frac{E_I^2 n}{2\mu c}.  \label{29}
\end{equation}
From Eq.~(\ref{25}) it moreover follows that $\tilde{n}_T$ becomes large and negative; setting $n_1=n+\Delta n$ with $\Delta n$ positive and small we get explicitly
\begin{equation}
\tilde{n}_T=-\frac{\kappa c^2}{\Delta n} <0. \label{30}
\end{equation}
Omitting the  $\Delta n$  corrections we thus obtain from Eq.~(\ref{28})
\begin{equation}
 \frac{E_T}{E_I}=\frac{\mu_1}{\mu}.
 \end{equation}
  The transmitted magnetic field follows from Eq.~(\ref{26}) as $H_T= E_Tn/(\mu_1c)$, and so
\begin{equation}
S_T=\frac{E_T^2n}{\mu c}=S_I. \label{32}
\end{equation}
This equality of the incident and transmitted Poynting vectors is somewhat surprising. The energy flow goes uninterrupted through the singular surface $z=0$ in the luminal limit; the reflected wave is zero.

Consider  the Minkowski momentum density $g_T^M$  in the transmitted region. As $ D_T={E_T\, n\tilde{n}_T/(\mu_1 c^2)}$, and $B_T=E_T\tilde{n}_T/c$, we get, letting $\mu_1 \rightarrow \mu$,
\begin{equation}
g_T^M=           \frac{1}{2}\Re {\bf(D\times B^*)}_z= \frac{E_T^2\, n{\tilde{n}}_T^2}{2\mu c^3}. \label{35}
\end{equation}
This is a large and positive quantity, caused by the large effective refractive index $\tilde{n}_T$.

Finally, the energy density in the transmitted region is
\begin{equation}
W_T= \frac{E_T^2n\tilde{n}_T}{2\mu c^2} <0.
\end{equation}
It satisfies the relation $S_T=W_T(c/\tilde{n}_T)$, similarly as for the incident wave.  The negativity of $W_T$ reflects the spacelike character of the Minkowski four-momentum. The positiveness of $S_T$ follows from both $W_T$ and $\tilde{n}_T$ being negative.

\bigskip

\noindent {\it 4. Connection with the  anomalous Doppler effect.} - The last issue we wish to emphasize, is the close connection that exists between the spacelike Minkowski four-momentum and the anomalous Doppler effect. This particular variant of the Doppler effect does not seem to be widely known; the theory of it was pioneered by  Frolov and Ginzburg \cite{ginzburg86}, and has later been followed up by others. We will here present an  exposition based upon   Ref.~\cite{brevik88}. It will now be convenient to adapt the common conventions in field theory: put $\hbar=c=1$, consider only the inertial frame where the medium, now assumed infinite, is at rest, and let a pointlike monopole detector  be moving in the positive $z$ direction with constant velocity ${\bf v}$. The refractive index is $n>1$, and we ignore dispersion.

As is usual, we replace thee electromagnetic field by a scalar field $\Phi$, assumed to response  to the presence of the medium (refractive index) in the same way as the electromagnetic field does. The detector has two energy levels, which in its rest system are called $E_{low}$  and $E_{top}$. In a transition, we let $E_{in}$ mean the initial level and $E_{f}$ the final one. The energy difference is
\begin{equation}
E=E_{f}-E_{in}. \label{36}
\end{equation}
If the detector becomes excited $(E_{in}=E_{low})$, then $E>0$. If it becomes deexcited $(E_{in}=E_{top})$, then $ E<0$.

Assume that the detector, of DeWitt type \cite{dewitt79}, is coupled to $\Phi$ via the interaction
\begin{equation}
L_{int}=\lambda m(\tau)\Phi[z(\tau)], \label{37}
\end{equation}
where $z(\tau)$ is the world line of the detector ($\tau$ is proper time), $m(\tau)$ is the monopole operator acting on the internal states, and $\lambda$ is  small coupling parameter.

Consider now the first order amplitude for the transition $E_{in}\rightarrow E_f$ with a simultaneous emission of a scalar field quantum of energy $\omega$ and momentum  ${\bf{k}}=n\omega {\hat{\bf k}}$,
\begin{equation}
A=i\lambda \langle E_f, 1_k|\int_{-\infty}^\infty m(\tau)\Phi[z(\tau)]
 d\tau |0, E_{in}\rangle. \label{38}
\end{equation}
By inserting $m(\tau)=\exp(iH_0t) m(0) \exp(-iH_0t)$ and expanding $\Phi $ in fundamental modes, we obtain
\begin{equation}
A=\frac{i\lambda \langle E_f|m(0)|E_{in}\rangle}{n\sqrt{16\pi^3\omega}}\int_{-\infty}^\infty d\tau \exp[i(E\tau -{\bf {k\cdot r}}(\tau)+\omega t(\tau)]]. \label{39}
\end{equation}
The probability for excitation or deexcitation with accompanying emission within the momentum volume element $d^3k=n^3\omega^2d\omega d\Omega$ is
\begin{equation}
d^3W =|A|^2 d^3k = \lambda^2|\langle E_f|m(0)|E_{in}\rangle |^2 d^3w, \label{40}
\end{equation}
with
\begin{equation}
d^3w = \frac{n\omega d\omega d\Omega}{16\pi^3}\left| \int_{-\infty}^\infty d\tau \exp [i[E\tau -kz \cos \theta+\omega t]]\right|^2, \label{41}
\end{equation}
a reduced probability independent of the details of the detector. Inserting now the time track of the detector, $z=vt=\gamma v\tau$ with $\gamma=(1-v^2)^{-1/2}$, we get
\begin{equation}
d^3w=\frac{n\omega d\omega d\Omega}{4\pi}|\delta [E+\gamma \omega (1-nv\cos\theta)]|^2,  \label{42}
\end{equation}
which shows that the emission angle $\theta_E$ is given by
\begin{equation}
\cos \theta_E= \frac{1}{nv}\left[ 1+\frac{E}{\gamma \omega}\right]. \label{43}
\end{equation}
Assume  that the velocity is superluminal, $nv>1$. The detector can be {\it excited} $(E>0)$, or {\it deexcited} $(E<0)$, by processes accompanied by {\it emission} of radiation. The detector alternates between the  levels $E_{low}$ and $E_{top}$. Let the emission angle for excitation be $\theta_E(\uparrow)$ and the corresponding angle for deexcitation be $\theta_E(\downarrow)$. From Eq.~(\ref{43}) we obtain
\begin{equation}
\cos \left\{\begin{array}{ll}
\theta_E(\uparrow)  \\
\theta_E(\downarrow)
\end{array}         \right\} =\frac{1}{nv}\left[1 \pm \frac{|E|}{\gamma \omega} \right]. \label{44}
\end{equation}
Thus $\theta_E(\uparrow)$ and $\theta_E(\downarrow)$ belong to  two different angular regions, the division line between them being the Cherenkov cone determined by $\cos \theta_C=1/(nv)$. The angle $\theta_E(\uparrow)$ corresponds to emission in the forward direction,
{\it inside} the Cherenkov cone. Quanta emitted in this region have negative energy. The negativeness of $\omega$ follows by a simple Lorentz transformation from the value in the  medium's rest frame.  We conclude that this is  precisely the spacelike character of the Minkowski four-momentum that surfaces again, although now formulated within scalar field theory. In electrodynamics, as mentioned, this is the anomalous Doppler effect.

Quanta emitted outside the Cherenkov cone, corresponding to $\theta_E(\downarrow)$, are associated with deexcitation of the detector, and positive scalar 'photon' energies.

To conclude this paper,  we have proposed an explanation of the experiment \cite{drori19} within classical electrodynamics, and have pointed out how the central ingredient - the spacelike Minkowski momentum - turns up also under other similar  circumstances involving superluminal velocities.

\end{document}